\documentclass[reprint, superscriptaddress, amsmath, amssymb, aps, prb]{revtex4-2}

\usepackage{amsmath}
\usepackage{amssymb}
\usepackage{indentfirst}
\usepackage{commath}
\usepackage{graphicx}
\usepackage[caption=false,position=bottom,labelfont={bf}]{subfig}
\usepackage{overpic}
\usepackage{dcolumn}
\usepackage{bm}
\usepackage{booktabs}
\usepackage{dsfont}
\usepackage{setspace}

\usepackage{xcolor}
\definecolor{myred}{rgb}{0.8,0.1,0.2}
\definecolor{myblue}{rgb}{0.1,0.2,0.6}

\usepackage[colorlinks,linktocpage,hypertexnames=false]{hyperref}
\hypersetup{
    colorlinks=true,
    linktoc=all,
    linkcolor={myred},
    citecolor={myblue},
    urlcolor={myblue},
}

\usepackage{xparse}
\ExplSyntaxOn
\NewDocumentCommand{\expt}{O{true} m G{} G{}}{
  \str_if_eq:nnTF {#1} {true}
    {
      \tl_if_empty:nTF {#3}
        {\left\langle #2 \right\rangle}
        {
          \tl_if_empty:nTF {#4}
            {\left\langle #2 \middle| #3 \right\rangle}
            {\left\langle #2 \middle| #3 \middle| #4 \right\rangle}
        }
    }
    {
      \tl_if_empty:nTF {#3}
        {\langle #2 \rangle}
        {
          \tl_if_empty:nTF {#4}
            {\langle #2 | #3 \rangle}
            {\langle #2 | #3 | #4 \rangle}
        }
    }
}
\ExplSyntaxOff


\begin{document}

\title{
Emergent magnetic pseudogap from phase fluctuations and hierarchy of scales in two-dimensional superconductors
}

\author{Xu-Cheng Wang}
\affiliation{State Key Laboratory of Surface Physics, Fudan University, Shanghai 200433, China}
\affiliation{Center for Field Theory and Particle Physics, Department of Physics, Fudan University, Shanghai 200433, China}

\author{Yang Qi}
\email{qiyang@fudan.edu.cn}
\affiliation{State Key Laboratory of Surface Physics, Fudan University, Shanghai 200433, China}
\affiliation{Center for Field Theory and Particle Physics, Department of Physics, Fudan University, Shanghai 200433, China}
\affiliation{Hefei National Laboratory, Hefei 230088, China}

\date{\today}

\begin{abstract}
Preformed pairs and phase fluctuations are believed to play a vital role in
predicting the charge pseudogap in the normal state of two-dimensional superconductors.
In this work, we extend this idea and further identify the emergent magnetic pseudogap
from pure phase fluctuations without invoking any competing order.
We examine the NMR relaxation rate $1/T_1T$
by evaluating the bubble contribution and leading-order vertex correction within perturbation theory.
It is found that the magnetic pseudogap, manifesting as a smooth suppression of $1/T_1T$ in the normal state,
is characterized by a temperature scale $T_\text{mPG}$ distinct from
the superconducting gap $\Delta_\text{SC}$ and transition temperature $T_c$.
The onset scales of both charge and magnetic pseudogap are dominated by
the competition of BKT correlation length $\xi(T)$ and BCS coherence length $\xi_\text{BCS}$,
i.e. the spatial extent of Cooper pair in the BCS theory.
Moreover, the vertex correction is shown to be irrelevant for $d$-wave pairing,
while it becomes prominent in $s$-wave systems
and drives a coherent enhancement of $1/T_1T$ at lower temperatures just above $T_c$.
We attribute this normal-state enhancement of $1/T_1T$
to the diverging coherence peak at the $s$-wave superconducting gap edge,
which shares the same spirit as the celebrated Hebel-Slichter peak in the BCS theory.
Analogous to the coherent Hebel-Slichter peak,
regularization by Fermi-liquid-like scatterings is important
and is characterized by a scattering length $\ell$.
The normal-state coherent enhancement of $1/T_1T$ is hence described by the competition of $\xi(T)$ and $\ell$,
through which the coherence scale $T_\text{coh}$ is determined.
In addition, the temperature scale $T_m$ associated with the dip of $1/T_1T$ is identified for $s$-wave systems.
As a result, the complete evolution of $1/T_1T$ is understood quantitatively in a unified picture
as the interplay among hierarchy of scales $\xi(T)$, $\xi_\text{BCS}$ and $\ell$.
Possible experimental and numerical observations of this emergent magnetic pseudogap
and coherent enhancement of $1/T_1T$ in the normal state are further discussed.
\end{abstract}

\maketitle

\section{Introduction}
Over the past decades, unconventional superconductivity remains an outstanding puzzle in condensed matter physics.
Empirically speaking, many unconventional superconductors (SCs) are effectively confined to two-dimensional (2D) space,
including cuprate materials~\cite{hashimoto2014energy,keimer2015quantum,sobota2021angle-resolved},
layered heavy fermion compounds~\cite{pfleiderer2009superconducting,zhou2013visualizing,
gyenis2018visualizing,wu2021revealing,smidman2023colloquium},
thin-film FeSe~\cite{song2011direct,kang2020preformed,faeth2021incoherent,jiang2023interplay}
and twisted multilayer graphene~\cite{cao2018unconventional,oh2021evidence,kim2022evidence}.
Dimensional effects dramatically shape the universality of superconducting transition,
giving rise to distinct behaviors as compared to conventional Bardeen-Cooper-Schrieffer (BCS) superconductors.

As stated by the Mermin-Wagner theorem,
no spontaneous breaking of continuous symmetry is allowed for systems with spatial dimension $d\leqslant2$.
Consequently, it is expected that the SC transition in 2D shall be governed by
the Berezinskii-Kosterlitz-Thouless (BKT) theory~\cite{kosterlitz1973ordering,kosterlitz1974critical}.
The BKT physics has been extensively
observed~\cite{xu2000vortex-like,
kang2020preformed,faeth2021incoherent,wang2023oscillating,weitzel2023sharpness}
in numerous quasi-2D superconducting compounds through a variety of experimental probes.
Among these studies, not only the BKT scaling of normal-state resistivity
was unambiguously measured~\cite{kang2020preformed,faeth2021incoherent},
but also the BKT fluctuations, i.e. vortex-antivortex unbinding, were detected~\cite{wang2023oscillating}.
It is further shown that taking into account the BKT phase fluctuations~\cite{emery1995importance} in the normal state
generally explains the emergence of pseudogap~\cite{franz1998phase,kwon1999effect,wang2023phase}
and Fermi arcs~\cite{berg2007evolution,han2010pseudogap,wang2023interplay},
which are both typically observed in doped cuprates,
and no competing order is required.

Apart from the photoemission spectroscopy, it is expected that
the BKT nature of SC transition in 2D shall also alter the general picture of spin response,
measured by advanced nuclear magnetic resonance (NMR) techniques~\cite{walstedt2018nmr}
through NMR shift, spin-lattice relaxation time $T_1$, spin-echo decay time $T_\text{2g}$, etc.
In particular, the measurement of spin-lattice relaxation time $T_1$
serves as a key probe of quasiparticle dynamics in superconductors.
One of the early confirmations of BCS theory for conventional SC is
its remarkable success in interpreting the $1/T_1$ peak immediately below the transition temperature $T_c$,
known as the Hebel-Slichter (HS) peak~\cite{hebel1957nuclear,hebel1959nuclear,coleman2015introduction}.
The presence of HS peak is a strong evidence of the coherent state with isotropic and nodeless gap.
As a result, the $T_1$ measurement offers a sensitive probe
in detecting $s$-wave or $d$-wave characteristics of the SC gap.
For instance, the calculations of $d$-wave $T_1$ characteristics below $T_c$
yielded one of the pioneering indicators of the $d$-wave singlet pairing in cuprates~\cite{monien1990spin,bulut1992weak}.

By convention, one may expect that the transition temperature $T_c$ and SC gap $\Delta_\text{SC}$
serve as the fundamental energy scales in BCS superconductors.
However, based on the experimental facts on quasi-2D SC compounds, e.g. cuprates,
an additional and distinct energy scale of \textit{magnetic pseudogap}~\cite{warren1989cu,
kambe1993nmr,yasuoka1997pseudo,goto1997phase,grafe2008nmr,hisashi2014strong}
$T^\ast>T_c$ emerges, which is widely recognized as the temperature
where $1/T_1T$ starts to drop from a nearly constant value at high $T$.
As suggested by the terminology, it is generally believed that
the collapse of $1/T_1T$ below $T^\ast$ indicates the rapid suppression of spin excitations.
To account for this distinct energy scale separated from $\Delta_\text{SC}$ and $T_c$,
a string of theoretical studies~\cite{imai1993low,
yasuoka1997pseudo,bankay1994single-spin,suter2000charge,uldry2005analysis}
attribute the collapse of $1/T_1T$ to either the onset of spin gap $\Delta_\text{spin}\simeq T^\ast$
or strong spin fluctuations in the paramagnetic phase.
Although discussed in several experiment-based proposals~\cite{janossy1997linear,
williams1997nmr,ishida1998pseudogap,raffa1998isotope,zheng2000superconducting,suter2000charge},
it remains however unclear whether the scale $T^\ast$ of magnetic pseudogap
can emerge entirely from superconducting fluctuations when competing orders are absent.

We hold a positive attitude to this possibility and an intuitive picture can be imagined as follows.
By ignoring the vertex correction and other complexities such as spin fluctuations,
the spin-lattice relaxation rate $1/T_1$ is approximately given by
the square of electronic density of states at the Fermi energy,
$1/T_1T\sim N(0)^2$~\cite{korringa1950nuclear,coleman2015introduction} for low temperatures $T$.
In this sense, the charge pseudogap manifesting in $N(0)$ and the magnetic pseudogap in $1/T_1T$ shall be closely related.
In our prior works~\cite{wang2023phase,wang2023interplay},
the universal connection between BKT phase fluctuations, charge pseudogap and Fermi arcs is established.
As a result, it is plausible to expect a generalized relation between phase fluctuations and magnetic pseudogap,
yet it remains unclear how the scale of magnetic pseudogap is explicitly determined through the BKT correlation length.
More importantly, the role of vertex correction is poorly understood.
And the fate of HS peak in the presence of phase fluctuations is unknown.

In this work, we address these queries based on our recently developed theoretical framework
on dealing with static phase fluctuations in the superconducting normal state.
By including both bubble graph and leading-order vertex correction within perturbation theory,
we calculated the spin-lattice relaxation rate $1/T_1T$ for 2D systems with both $s$-wave and $d$-wave pairing.
Our core findings are illustrated in Fig.~\ref{fig:fig2}.
It is found that for both $s$-wave and $d$-wave systems,
a prominent magnetic pseudogap emerges solely from the phase fluctuations,
and its evolution, analogous to the onset of charge pseudogap, is dominated by the competition of
BKT correlation length $\xi(T)$ and BCS coherence length $\xi_\text{BCS}$,
with $\xi_\text{BCS}$ characterizing the spatial extent of Cooper pair.
The emergent energy scale $T_\text{mPG}$ of magnetic pseudogap determined via $\xi(T_\text{mPG})\sim\xi_\text{BCS}$
does not correspond to the onset of spin fluctuations and is also clearly distinct from the SC gap.
Furthermore, as $T_c$ is approached in the normal state,
we observed an unexpected enhancement of $1/T_1T$ in $s$-wave SCs,
which is attributed to the diverging vertex correction at $T_c$.
In contrast, this enhancement does not occur for $d$-wave systems
since the vertex correction is asymptotically suppressed at $T_c$ due to the nodal form factor.
We note that the divergence of $s$-wave vertex correction at $T_c$
stems from the infinitely sharp coherence peak formed at the edge of SC gap,
sharing an analogous spirit with the coherent HS peak,
and hence can be regularized by introducing additional background scatterings $\Gamma_0$.
As compared to the HS peak, the coherent enhancement of $1/T_1T$ here develops in the SC normal state
and primarily arises from the vertex correction;
however, the HS peak manifests just below $T_c$ and can be well understood
on the mean-field level by considering the bubble graph alone.
We subsequently identified this coherent enhancement of $1/T_1T$
as the interplay of $\xi(T)$ and the characteristic length of background scatterings $\ell=v_F/\Gamma_0$,
and determined the associated coherence energy scale $T_\text{coh}$.
In addition, the dip of $1/T_1T$ in the normal state with characteristic scale $T_m$ was further examined for $s$-wave SCs.
Possible verifications of our findings in quasi-2D SC materials and correlated lattice models were also discussed.

The remaining contents are organized as follows.
In Sec.~\ref{sec:pdsc}, we first introduce the theoretical framework used to deal with static phase fluctuations
and highlight its applications on predicting charge pseudogap and Fermi arcs.
Then, in Sec.~\ref{sec:kubo}, we briefly review the Kubo formula,
which connects the spin-lattice relaxation rate to spin susceptibility through the fluctuation-dissipation theorem.
We apply our theory in Sec.~\ref{sec:susceptibility} to calculate the spin susceptibility of phase-fluctuating SCs,
taking into account both bubble contribution and leading-order vertex correction.
Our central theoretical findings are summarized in Sec.~\ref{sec:results}.
In Sec.~\ref{sec:mpg-scale}, the emergent scale $\xi(T_\text{mPG})$ of magnetic pseudogap
is determined through analyzing the bubble contribution,
and in Sec.~\ref{sec:coh-peak}, we delve into the role of vertex correction
and explore in detail the scales $\xi(T_\text{coh})$ and $\xi(T_m)$ concerning the coherent enhancement of $1/T_1T$.
Finally, we leave the conclusion and relevant discussions in Sec.~\ref{sec:conclusion}.

\section{Theoretical formalism}

\subsection{Phase-fluctuating superconductivity}\label{sec:pdsc}
To begin with, we briefly review our approach to dealing with 2D phase-fluctuating superconductivity.
For a detailed discussion, readers may refer to Refs.~\cite{wang2023phase,wang2023interplay}.
To describe 2D superconductors in the normal state with short-ranged superconducting correlations,
we consider a minimal phenomenological Hamiltonian $H=H_0+V$ with
\begin{subequations}\begin{align}
    H_0 &= \sum_\sigma \int\mathrm{d}^2r\ \psi^\dag_\sigma(r) \left(-\frac{\nabla^2}{2m}-\mu\right) \psi_\sigma(r),\label{eq:h0}\\[5pt]
    V &= \int\mathrm{d}^2r\ \mathrm{d}^2s\ \Delta(r,s)\psi^\dag_\uparrow(r+s/2)\psi^\dag_\downarrow(r-s/2) + \text{h.c.},\label{eq:v}
\end{align}\end{subequations}
where $\psi_\sigma(r)/\psi^\dag_\sigma(r)$ annihilates/creates an electron at site $r$ with spin $\sigma$,
and $\Delta(r,s)$ represent the static, spatially fluctuating superconducting order parameters with a general pairing symmetry.
The pairing symmetry is explicitly revealed after one performs
the Fourier transform with respect to the internal displacement $s$ of the Cooper pair,
$\int\mathrm{d}^2s\ e^{-ips}\Delta(r,s)=\Delta(r)\varphi(p)$,
where $\varphi(p)$ denotes the pairing form factor,
e.g. $\varphi(p)=1$ for the $s$-wave pairing and $\varphi(p)=(p_x^2-p_y^2)/p^2$ for the $d$-wave pairing.

By adopting the time-independent Hamiltonian in Eqs.~\eqref{eq:h0}\eqref{eq:v},
we have omitted the quantum fluctuations in the imaginary-time direction.
This is well justified for finite temperatures close to the BKT point.
At finite temperatures, the temporal correlation length is bounded by the inverse temperature $\beta$,
while the spatial correlation $\xi$ diverges at the BKT point such that $\beta\ll \xi/v_F$.
Hence, for a low-energy theory satisfying $v_F\beta<a<\xi$ with certain spatial cutoff $a$,
the temporal fluctuations are effectively integrated out.

Given certain configuration of $\{\Delta(r,s)\}$,
the model in Eqs.~\eqref{eq:h0}\eqref{eq:v} is exactly solvable since it is quadratic.
In order to take into account the spatial fluctuations of $\Delta(r,s)$,
we introduce the disorder averaging conditions
\begin{subequations}\begin{align}
    \overline{\Delta(r)} &= \overline{\Delta^\ast(r)} = 0,\label{eq:disorder-averaging-1}\\[5pt]
    \overline{\Delta(r)\Delta^\ast(r')} &= \abs{\Delta_0}^2\ g\left(\vert r-r'\vert/\xi\right),\label{eq:disorder-averaging-2}
\end{align}\end{subequations}
where $g(x)$ is a dimensionless function
which decays exponentially from $g(0)=1$ to 0 for $x\gg1$.
In the spirit of BKT transition, we consider here the phase fluctuations alone
but maintain a uniform, temperature-independent pairing amplitude $\Delta_0$.
As implied in Eq.~\eqref{eq:disorder-averaging-1},
no superconducting order survives locally after averaging over the fluctuating pairing configurations,
which is as expected for the normal state.
The BKT superconducting correlation length $\xi$ enters
through the non-vanishing two-point correlation in Eq.~\eqref{eq:disorder-averaging-2}.
At the transition temperature $T_c$, the correlation length $\xi$ formally diverges,
while in the normal state it follows the BKT scaling relation as
$\xi(T)\sim \exp(bt_r^{-1/2})$~\cite{kosterlitz1973ordering,kosterlitz1974critical}
with the reduced temperature $t_r=(T-T_c)/T_c$.

\begin{figure}[htbp]
    \centering
    \includegraphics[width=.75\columnwidth]{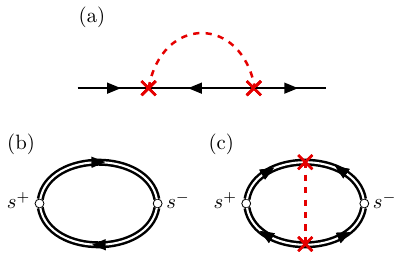}
    \caption{%
        (a) Leading-order correction to the electronic Green's function,
            from which the self-energy Eq.~\eqref{eq:self-energy} is deduced.
            The red dashed line denotes the average over fluctuating pairing correlations.
        (b) Bubble graph to the spin susceptibility $\chi_{-+}(q,i\omega_n)$.
            The doubled line represents the interacting Green's function $G^{-1}=[G^{(0)}]^{-1}-\Sigma$.
        (c) Leading-order vertex correction to $\chi_{-+}(q,i\omega_n)$.
    }
    \label{fig:fig1}
\end{figure}

Assuming that the superconducting pairing $\Delta_0$ is weak
as compared to the characteristic energy of electrons, e.g. Fermi energy $E_F$ for the continuum model,
we can regard Eq.~\eqref{eq:v} as a perturbation.
Combining perturbation theory with the averaging conditions in Eqs.~\eqref{eq:disorder-averaging-1}\eqref{eq:disorder-averaging-2}
then enables us to calculate various physical observables.
The first intriguing aspect is the single-particle properties of this theory.
In the weak-fluctuating regime where $k_F\xi\gg1$,
the leading-order self-energy~\cite{wang2023phase,wang2023interplay}, as seen in Fig.~\ref{fig:fig1}(a), is given by
\begin{equation}\label{eq:self-energy}
    \Sigma(k,\omega) = \frac{\abs{\Delta_k}^2}{\omega+\xi_k+2i\Gamma_k},
\end{equation}
with the superconducting gap function $\Delta_k=\Delta_0\varphi(k)$.
The \textit{pair-breaking rate} $\Gamma_k=v_k/2\xi$
denotes the inverse lifetime of Cooper pairs in the presence of normal-state phase fluctuations,
which directly contributes to the finite imaginary component of the self-energy.
$\xi_k=k^2/2m-\mu$ is the energy dispersion and $v_k=\abs{k}/m$ the velocity of free electrons as usual.
The electron retarded Green's function $G(k,\omega)$ is related to the self-energy $\Sigma(k,\omega)$
through the Dyson series,
$G(k,\omega)^{-1}=[G^{(0)}(k,\omega)]^{-1}-\Sigma(k,\omega)$,
where $G^{(0)}(k,\omega)=1/(\omega-\xi_k+i\eta)$ is the bare retarded Green's function
with $\eta\to0^+$ an infinitesimal, positive quantity.
Then the single-particle spectral function $A(k,\omega)=-2\text{Im}[G(k,\omega)]$
is simply the imaginary component of $G(k,\omega)$.
It is shown that as temperature $T$ increases from $T_c$,
the phase fluctuations proliferate
and the correlation length $\xi$ rapidly becomes short-ranged,
resulting in the growth of $\Gamma_k$ and the smearing of coherent peaks in $A(k,\omega)$.
This intuitive picture predicts the evolution of charge pseudogap in 2D phase-fluctuating superconductors,
i.e. the gradual filling of SC gap in $A(k,\omega)$.
For $s$-wave pairing, a striking conclusion is that the evolution of charge pseudogap in $A(k,\omega)$ is dominated by
the BKT correlation length $\xi(T)$ and the BCS coherence length $\xi_\text{BCS}=v_F/\pi\Delta_0$,
as shown in Fig.~\ref{fig:fig2}(c).
$\xi_\text{BCS}$ characterizes the spatial size of Cooper pair in the weak-coupling BCS theory~\cite{annett2003superconductivity} with $v_F$ the Fermi velocity.
The pseudogap at the Fermi energy completely closes when~\cite{wang2023phase}
\begin{equation}\label{eq:cpg-scale}
    \frac{\xi(T_\text{cPG})}{\pi\xi_\text{BCS}} = \frac{1}{\sqrt{2}},
\end{equation}
from which the temperature scale of charge pseudogap $T_\text{cPG}$ is deduced.
For $d$-wave pairing, it is revealed in Ref.~\cite{wang2023interplay} that
the anisotropy of nodal gap further leads to Fermi arcs following the same spirit,
as illustrated in Fig.~\ref{fig:fig2}(d).

\subsection{Kubo formula for the transverse spin-lattice relaxation rate}\label{sec:kubo}
Within the framework of linear response theory,
the \textit{transverse spin-lattice relaxation rate} $1/T_1$,
also referred to as the NMR relaxation rate, is defined in terms of the spin structure factor as follows
~\cite{sandvik1995nmr,sato2009nmr,dupont2016temperature,nasu2021spin}
\begin{equation}\label{eq:spin-relaxation-rate}
    1/T_1 = \frac{1}{V} \sum_{q,\alpha=x,y} \abs{A^\alpha_\text{hf}(q)}^2 S_{\alpha\alpha}(q,\omega\to0).
\end{equation}
$A^\alpha_\text{hf}(q)$ denotes the hyperfine coupling between the nucleus and the electron spin, 
and $\alpha$ enumerates the two axes perpendicular to the direction of the external field.
For finite-temperature system with vanishing local moment,
the \textit{dynamic spin structure factor}~\cite{sachdev2011quantum}
is defined through the spectral decomposition,
{\small
\begin{equation}\begin{aligned}\label{eq:spectral-decomposition}
    &S_{\alpha\alpha'}(q,\omega)
     = \frac{1}{V} \int_{-\infty}^{+\infty} \mathrm{d}t\ e^{i\omega t} \expt{s^\alpha_q(t)s^{\alpha'}_{-q}(0)}\\[5pt]
    &= \frac{2\pi}{V\mathcal{Z}}\sum_{mm'} \expt[false]{m}{s^\alpha_q}{m'} \expt[false]{m'}{s^{\alpha'}_{-q}}{m}\ e^{-\beta E_m}\ \delta\left(\omega+E_m-E_{m'}\right),
\end{aligned}\end{equation}
}\noindent
where $O(t)=e^{iHt}O(0)e^{-iHt}$ for time-dependent operator
and $\expt{\cdots}$ stands for the average over ensembles in the thermal equilibrium.
$\mathcal{Z}=\text{Tr}\left[e^{-\beta H}\right]$ is the partition function at temperature $1/\beta$.
The local spin operator is defined as
$s^\alpha_r=c^\dag_{r,\sigma} [S^\alpha]_{\sigma\sigma'} c_{r,\sigma'}$,
where $S^\alpha$ denote the Pauli spin matrices.
The summations over spin indices $\sigma$ and $\sigma'$ are implicitly assumed.
Then the spin operators in momentum space take the form of
$
    s^\alpha_q = \sum_r e^{-iqr} s^\alpha_r = \sum_p c^\dag_{p,\sigma} [S^\alpha]_{\sigma\sigma'} c_{p+q,\sigma'}
$.
As indicated in Eq.~\eqref{eq:spin-relaxation-rate},
the spin-lattice relaxation rate is related to the static spin structure factor $S_{\alpha\alpha}(\omega\to 0)$,
because the NMR frequency always serves as the minimal energy scale of the system
as compared to the Fermi energy $E_F$, superconducting gap $\Delta_0$ or the thermal energy scale $k_BT$.

Throughout this work, we assume an isotropic and local hyperfine coupling
(due to the locality of the nucleus-electron interaction)
such that $A^\alpha_\text{hf}(q)=A_\text{hf}$.
It is practically convenient to express Eq.~\eqref{eq:spin-relaxation-rate}
in terms of spin raising and lowering operators $s^{\pm}=s^x\pm is^y$, which leads to
\begin{equation}\begin{aligned}\label{eq:spin-relaxation-rate-2}
    1/T_1
    &= \frac{1}{2V} \abs{A_\text{hf}}^2 \sum_{q} \left[S_{-+}(q,\omega\to0) + S_{+-}(q,\omega\to0)\right]\\
    &= \frac{1}{V} \abs{A_\text{hf}}^2 \sum_{q} S_{-+}(q,\omega\to0).
\end{aligned}\end{equation}
In the second step, we have noticed that $S_{-+}(q,\omega\to0) = S_{+-}(-q,\omega\to0)$
by explicitly checking their spectral decompositions.
Further we define the \textit{dynamic transverse spin susceptibility} $\chi_{-+}(q,\omega)$,
the imaginary part of which is related to the structure factor through the \textit{fluctuation-dissipation theorem} as
$
    \text{Im}\left[\chi_{-+}(q,\omega)\right] = \frac{1-e^{-\beta\omega}}{2} S_{-+}(q,\omega)
$.
This will eventually give
\begin{equation}\label{eq:spin-relaxation-rate-kubo}
    1/T_1T = 2\abs{A_\text{hf}}^2 \frac{1}{V}\sum_q \lim_{\omega\to 0^+} \frac{\text{Im}\left[\chi_{-+}(q,\omega)\right]}{\omega}.
\end{equation}
Eq.~\eqref{eq:spin-relaxation-rate-kubo} is known as the Kubo formula for the transverse spin-lattice relaxation rate,
and serves as the foundation of our theoretical analyses.
For simplicity, we have set $\abs{A_\text{hf}}=1$ for all the calculations presented in this work.

\subsection{Transverse spin susceptibility for phase-fluctuating superconductors}\label{sec:susceptibility}
For our purpose of calculating the NMR relaxation rate,
in this section we evaluate the spin response function
$
    \chi_{-+}(q,i\omega_n) = \frac{1}{V} \int_0^\beta\mathrm{d}\tau e^{i\omega_n\tau} \expt{T_\tau s^-_{q}(\tau)s^+_{-q}(0)}
$
for phase-fluctuating SCs
with Matsubara frequency $\omega_n=2n\pi/\beta$, $n\in\mathbb{Z}$.
Expanding $\chi_{-+}(q,i\omega_n)$ with respect to $\abs{\Delta_0}$,
the leading-order contribution is illustrated as the bubble graph in Fig.~\ref{fig:fig1}(b)
and given by Eq.~\eqref{eq:bubble-graph}.
We denote the interacting Matsubara Green's function $G(k,i\nu_n)$
as the doubled lines in Fig.~\ref{fig:fig1}(b)(c).
$G(k,i\nu_n)$ is inferred from the self-energy $\Sigma(k,i\nu_n)$, Fig.~\ref{fig:fig1}(a), through Dyson series,
where the effects of phase fluctuations have been included.
$\nu_n=(2n+1)\pi/\beta$ with $n\in\mathbb{Z}$ is the fermionic Matsubara frequency.

\begin{widetext}\begin{align}
    &\qquad\qquad\qquad\quad
    \chi^{\text{bd}}_{-+}(q,i\omega_n)
    = -\frac{1}{V\beta} \sum_{p,i\nu_n} G_\downarrow(p,i\nu_n) G_\uparrow(p+q,i\omega_n+i\nu_n),\label{eq:bubble-graph}\\[5pt]
    \chi^\text{vc}_{-+}(q,i\omega_n)
    &= -\frac{1}{V^3\beta}\sum_{p_1,p_2,i\nu_n} \overline{\Delta_{p_1,p_2}\Delta^\ast_{p_1+q,p_2-q}}
    \ G_\downarrow(p_1,i\nu_n) G_\uparrow(p_2,-i\nu_n) G_\uparrow(p_1+q,i\nu_n+i\omega_n) G_\downarrow(p_2-q,-i\nu_n-i\omega_n).\label{eq:vertex-correction}
\end{align}\end{widetext}

To the sub-leading $\abs{\Delta_0}^2$ order, the vertex correction enters
as illustrated in Fig.~\ref{fig:fig1}(c) and given by Eq.~\eqref{eq:vertex-correction}.
Without loss of generality and to facilitate our calculation~\cite{wang2023interplay},
we consider a Gaussian pairing correlation
$
    g\left(\vert r-r'\vert/\xi\right)=e^{-\abs{r-r'}^2/2\xi^2}
$
in Eq.~\eqref{eq:disorder-averaging-2}.
As a result, the averaged pairing correlation in the momentum space in Eq.~\eqref{eq:vertex-correction}
can be worked out explicitly as
\begin{equation}\begin{aligned}
    &\overline{\Delta_{p_1p_2}\Delta^\ast_{p_3p_4}}
    = V \delta^{(2)}(p_1+p_2-p_3-p_4)\\[5pt]
    &\times \abs{\Delta_0}^2 2\pi\xi^2 e^{-(p_1+p_2)^2\xi^2/2}
    \varphi\left(\frac{p_1-p_2}{2}\right) \varphi^\ast\left(\frac{p_3-p_4}{2}\right),
\end{aligned}\end{equation}
where the superconducting order parameter in the momentum space reads
\begin{equation}\begin{aligned}
    \Delta_{p_1p_2}
    &= \int \mathrm{d}^2r\ \mathrm{d}^2s\ \Delta(r,s)\ e^{-ip_1(r+s/2)-ip_2(r-s/2)}\\
    &= \int \mathrm{d}^2r\ \Delta(r)\ e^{-i(p_1+p_2)r} \varphi\left(\frac{p_1-p_2}{2}\right).
\end{aligned}\end{equation}

It is notable that in our theory,
the anomalous propagation of electron is forbidden in the normal state
such that the electrons can only propagate normally in Figs.~\ref{fig:fig1}(b)(c).
Substituting Eqs.~\eqref{eq:bubble-graph}\eqref{eq:vertex-correction} into Eq.~\eqref{eq:spin-relaxation-rate-kubo}
then provides us with the estimation of $1/T_1T$ to the $\abs{\Delta_0}^2$ order.
We will discuss the results in detail in Sec.~\ref{sec:results}.

\section{Results}\label{sec:results}
\begin{figure*}[htbp]
    \centering
    \hspace*{-.7cm}
    \includegraphics[width=.9\linewidth]{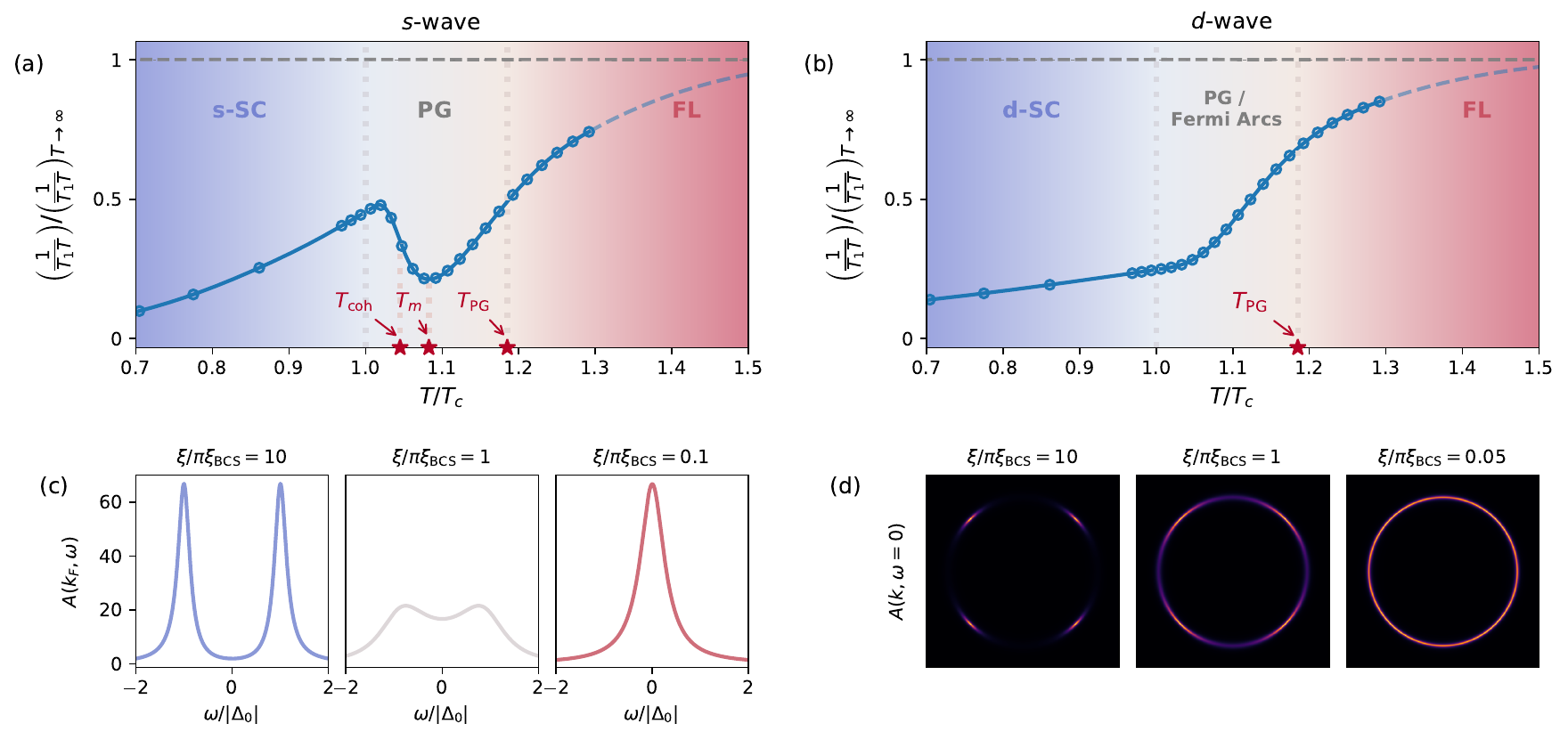}\\
    \vspace*{-.2cm}
    \includegraphics[width=.9\linewidth]{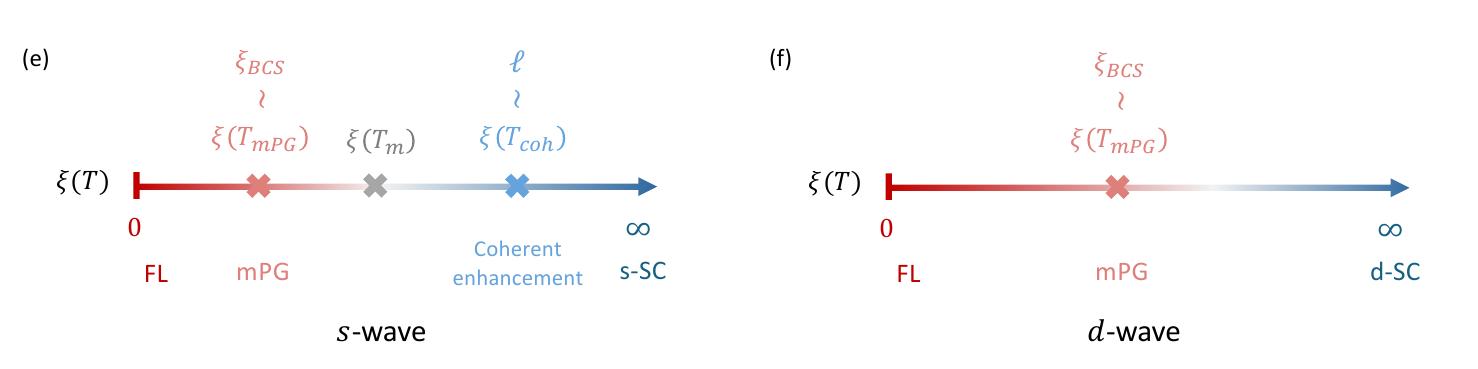}
    \caption{%
        (a)(b) Spin-lattice relaxation rate $1/T_1T$ with respect to temperature $T$,
               including bubble-graph contribution and leading-order vertex correction.
               The correlation length is given by the BKT relation $\xi(T)=\xi_0 \exp(bt_r^{-1/2})$ in the normal state,
               and set to infinity for $T$ below $T_c$.
               Parameters are taken as follows:
               $m=1$, $k_F=2.2$, $\beta_c=7.75$, $\Delta_0=0.6$, $\Gamma_0=0.1$, $b=1.4$, $\xi_0=0.1$.
               The emergent scale of magnetic pseudogap $\xi(T_\text{mPG})$ is marked out.
               For an $s$-wave system, the scales of coherent enhancement $\xi(T_\text{coh})$ and $1/T_1T$ dip $\xi(T_m)$ are additionally identified.
        (c)(d) Evolution of (c) charge pseudogap and (d) Fermi arcs in $d$-wave system
               as a competition of $\xi$ and $\xi_\text{BCS}=v_F/\pi\abs{\Delta_0}$.
        (e)(f) Hierarchy of scales governing the evolution of $1/T_1T$.
               $T_\text{mPG}$, $T_\text{coh}$ are determined through
               $\xi(T_\text{mPG})\sim\xi_\text{BCS}$ and $\xi(T_\text{coh})\sim\ell=v_F/\Gamma_0$ respectively.
    }
    \label{fig:fig2}
\end{figure*}

The bubble-diagram contribution and leading-order vertex correction to $1/T_1T$ are shown respectively
in Eq.~\eqref{eq:nmr-bd} and Eqs.~\eqref{eq:nmr-vc}\eqref{eq:I},
where $n_F$ is the Fermi-Dirac distribution.
To finally arrive at these organized results,
it is convenient to start from Eqs.~\eqref{eq:bubble-graph}\eqref{eq:vertex-correction}
and first express the Green's functions therein in the Lehmann spectral representation,
$G(k,i\nu_n)=\int\frac{\mathrm{d}\varepsilon}{2\pi}\frac{A(k,\varepsilon)}{i\nu_n-\varepsilon}$.
To proceed, one completes the Matsubara sums over the free propagators, which is a standard yet tedious procedure.
The NMR relaxation rate then follows explicitly from
Eqs.~\eqref{eq:spin-relaxation-rate-kubo}\eqref{eq:bubble-graph}\eqref{eq:vertex-correction}.

\begin{widetext}
    \begin{equation}\begin{aligned}\label{eq:nmr-bd}
        \left(1/T_1T\right)^\text{bd}
        &= \abs{A_\text{hf}}^2 \int\frac{\mathrm{d}^2p}{(2\pi)^2} \int\frac{\mathrm{d}^2q}{(2\pi)^2} \int\frac{\mathrm{d}\varepsilon}{2\pi}
           \left(-\frac{\mathrm{d}n_F}{\mathrm{d}\varepsilon}\right) A_\downarrow(p,\varepsilon) A_\uparrow(p+q,\varepsilon)\\[5pt]
        &= \abs{A_\text{hf}}^2 \int\frac{\mathrm{d}\varepsilon}{2\pi}
           \left(-\frac{\mathrm{d}n_F}{\mathrm{d}\varepsilon}\right) N_\downarrow(\varepsilon) N_\uparrow(\varepsilon),
    \end{aligned}\end{equation}
\end{widetext}
\begin{widetext}
    \begin{subequations}\begin{align}
        &\qquad\quad \left(1/T_1T\right)^\text{vc}
        = \abs{A_\text{hf}}^2 \int\frac{\mathrm{d}^2p}{(2\pi)^2} \int\frac{\mathrm{d}\varepsilon}{2\pi}
          \left(-\frac{\mathrm{d}n_F}{\mathrm{d}\varepsilon}\right)\ \abs{\Delta_0}^2 2\pi\xi^2\ e^{-p^2\xi^2/2}
          \times I_{\downarrow\uparrow}(p,\varepsilon)\ I^\ast_{\uparrow\downarrow}(p,\varepsilon),\label{eq:nmr-vc}\\[8pt]
        &I_{\sigma\bar{\sigma}}(p,\varepsilon)
        = \int\frac{\mathrm{d}^2\tilde{p}}{(2\pi)^2} \int\frac{\mathrm{d}\tilde{\varepsilon}}{2\pi}
          \ \frac{1}{\tilde{\varepsilon}}\ \varphi\left(\frac{p}{2}-\tilde{p}\right)
          \left[A_\sigma(p-\tilde{p},\varepsilon) A_{\bar{\sigma}}(\tilde{p},\tilde{\varepsilon}-\varepsilon) - A_\sigma(p-\tilde{p},\tilde{\varepsilon}+\varepsilon) A_{\bar{\sigma}}(\tilde{p},-\varepsilon)\right].\label{eq:I}
    \end{align}\end{subequations}
\end{widetext}

We first comment on the overall structure of the results.
The bubble contribution Eq.~\eqref{eq:nmr-bd} and the vertex correction Eq.~\eqref{eq:nmr-vc}
are positive definite due to the obvious modular square structure.
(Both $A_\sigma(p,\varepsilon)$, $N_\sigma(\varepsilon)$ and $I_{\sigma\bar{\sigma}}(p,\varepsilon)$ are spin-invariant.)
Intuitively, this modular square structure originates from the electrons propagating
along two time-reversal related paths connecting $s^+$ and $s^-$,
as illustrated in Figs.~\ref{fig:fig1}(b)(c),
and is guaranteed by the optical theorem~\cite{peskin1995introduction}.
Furthermore, the vertex correction for $d$-wave pairing vanishes exactly
in the weak-fluctuating regime with large correlation length $k_F\xi\gg1$.
In this case, the integral over $p$ in Eq.~\eqref{eq:nmr-vc} can be replaced with a $\delta$-function
such that the remaining integrand is directly proportional to $I_{\sigma\bar{\sigma}}(0,\varepsilon)$.
By identifying $\varphi$ in Eq.~\eqref{eq:I} as the $d$-wave form factor,
the angle integral over $\tilde{p}$ in $I_{\sigma\bar{\sigma}}(0,\varepsilon)$ is exactly cancelled
and hence the vertex correction is fully suppressed.

It is also notable that for both $s$-wave and $d$-wave pairing,
the bubble contribution Eq.~\eqref{eq:nmr-bd} formally diverges as $T_c$ is approached in the normal state.
This is because at $T_c$ and in the SC phase,
the correlation length $\xi$ becomes infinitely long-ranged,
and the self-energy Eq.~\eqref{eq:self-energy} and Green's function reduce to those of the BCS theory.
In this case, the integral in Eq.~\eqref{eq:nmr-bd} is sensitive to
the sharp coherence peak around $\abs{\omega}\simeq\abs{\Delta_0}$ in the BCS local density of states (LDOS).
We recall that for the $s$-wave BCS theory,
the LDOS $N_s(\omega)\simeq N(0)\abs{\omega}/\sqrt{\omega^2-\Delta_0^2}$,
which brings logarithmic divergence to $(1/T_1T)^\text{bd}$.
For $d$-wave pairing, $N_d(\omega)\simeq \frac{2}{\pi}N(0) K(\Delta_0^2/\omega^2)$
where $K(z)$ is known as the complete elliptic integral of the first kind.
Recalling the limiting behavior $\lim_{z\to1} [K(z)-\ln(4/\sqrt{1-z})] = 0$~\cite{abramowitz1968handbook,byrd1971handbook},
we thus have $N_d(\omega)\simeq \frac{2}{\pi}N(0)\ln[4N_s(\omega)/N(0)]$ near the gap edge,
which also diverges at $\abs{\omega}\simeq\abs{\Delta_0}$ but with a much slower rate of divergence as compared to $N_s$.
Moreover, in addition to the bubble contribution,
it is verified that the $s$-wave vertex correction in Eq.~\eqref{eq:nmr-vc}
is also strongly influenced by the sharp coherence peak of LDOS
with an even higher divergence rate than the bubble graph,
since it involves a quartic product of spectral functions.

In order to regulate this divergence arising from coherence peak,
we introduce a background broadening to the single-particle spectrum.
This approach is analogous to the treatment of the HS peak in BCS theory~\cite{coleman2015introduction},
where disorder-induced scattering plays a crucial role in regularizing the divergence therein.
As compared to Eq.~\eqref{eq:self-energy}, we consider the following modified self-energy,
\begin{equation}
    \Sigma(k,\omega) = \frac{\abs{\Delta_k}^2}{\omega+\xi_k+2i\Gamma_k} - i\Gamma_0.
\end{equation}
A small and negative imaginary part $-i\Gamma_0$ is added
such that $\text{Im}\left[\Sigma\right]$ now is finite-valued even when $\xi$ diverges with $\Gamma_k\to0^+$.
From a realistic perspective, $\Gamma_0$ can stem from
non-magnetic disorders distributed in the material,
interaction effects among electrons within Landau's Fermi liquid framework,
or amplitude fluctuations of SC in the normal state.
To further simplify the problem, we ignore the temperature dependence of $\Gamma_0$
within certain temperature window near $T_c$,
and we define the associated length scale of scattering $\ell=v_F/\Gamma_0$,
which is later shown to describe the coherent enhancement of $1/T_1T$ together with $\xi(T)$.
For system with weak background scattering, we expect $\ell\gg\xi_\text{BCS}$.
Under these assumptions, Eqs.~\eqref{eq:nmr-bd}\eqref{eq:nmr-vc}\eqref{eq:I} can be evaluated without ambiguity
through solving the high-dimensional integrals with advanced numerical techniques.

\begin{figure}[htbp]
    \centering\hspace*{-.7cm}
    \includegraphics[width=.98\columnwidth]{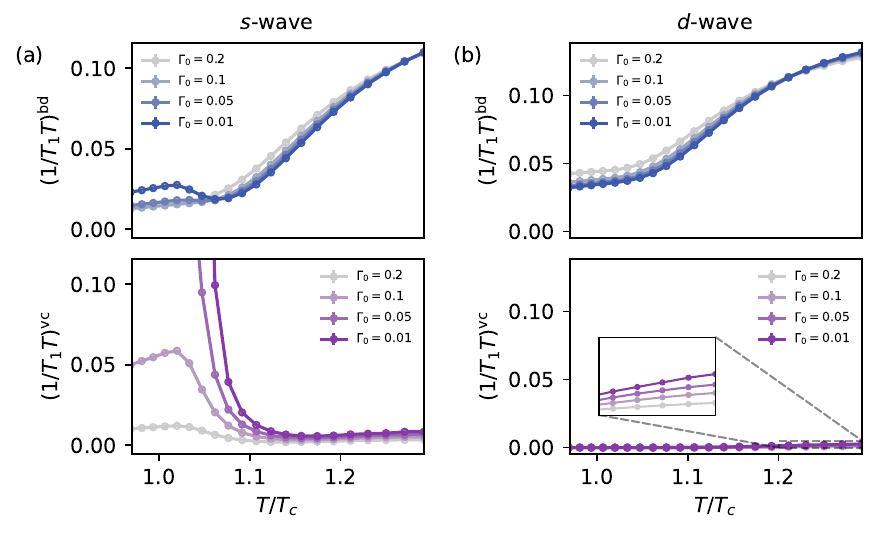}
    \caption{%
        Bubble-graph contribution $(1/T_1T)^\text{bd}$ and leading-order vertex correction $(1/T_1T)^\text{vc}$
        with respect to temperature $T$ for (a) $s$-wave and (b) $d$-wave systems.
        The specific model parameters are identical to those adopted in Fig.~\ref{fig:fig2},
        except that the background scattering strength $\Gamma_0$ is varied.
    }
    \label{fig:fig3}
\end{figure}

The central results of this work are summarized in Fig.~\ref{fig:fig2}
and we show separately the bubble contribution and vertex correction in Fig.~\ref{fig:fig3}.
One shall expect that at sufficiently high temperatures,
the $1/T_1T$ exhibits the anticipated Fermi liquid (FL) behavior
that a nearly constant value is reached providing $\beta E_F\ll1$.
As the temperature $T$ decreases, the BKT correlation length $\xi(T)$ develops,
and the spin-lattice relaxation rate $1/T_1T$ rapidly drops from the FL value
far before $T_c$ is reached for both $s$-wave and $d$-wave SCs.
This marks the onset of magnetic pseudogap in the normal state,
accompanied by the opening of single-particle charge pseudogap as shown in Fig.~\ref{fig:fig2}(c),
which is regarded as a precursor to the SC gap.
For $d$-wave pairing, the anisotropy of the nodal gap further leads to the Fermi arcs in Fig.~\ref{fig:fig2}(d).
It is found that similar to the charge pseudogap described by Eq.~\eqref{eq:cpg-scale},
the evolution of magnetic pseudogap is dominated by two length scales,
i.e. the BKT correlation length $\xi(T)$ and the spatial size of Cooper pair $\xi_\text{BCS}=v_F/\pi\Delta_0$.
We characterize the temperature scale $T_\text{mPG}$ of magnetic pseudogap as the temperature where $1/T_1T$ is most rapidly suppressed,
and identify
\begin{equation}
    T_\text{mPG}:\quad \xi(T_\text{mPG}) \sim \xi_\text{BCS}.
\end{equation}
That is, $T_\text{mPG}$ can be understood as the temperature at which $\xi(T)$ becomes comparable with $\xi_\text{BCS}$.
In Sec.~\ref{sec:mpg-scale}, we will explore in detail the quantitative relation between $\xi(T_\text{mPG})$ and $\xi_\text{BCS}$
by examining the bubble contribution.

As $T$ proceeds to decrease,
for $s$-wave SC and with moderate scattering length $\ell=v_F/\Gamma_0$,
$1/T_1T$ undergoes an unexpected enhancement in the normal state,
while for $d$-wave SC such $1/T_1T$ peak does not appear with reasonably large $\ell$.
This NMR peak manifests as an immediate consequence of the explosion of $s$-wave vertex correction near $T_c$,
which is illustrated in Fig.~\ref{fig:fig3}(a).
We attribute the explosion of $s$-wave vertex correction to
the development of sharp coherence peak in the electronic single-particle spectrum,
and regularize it through incorporating the finite background scattering length $\ell$.
Therefore, the coherent enhancement of $1/T_1T$ shares the same origin with the coherent HS peak in the $s$-wave BCS theory.
It is found that the temperature scale $T_\text{coh}$ associated with this coherent enhancement is determined through
the competition of BKT correlation length $\xi(T)$ and background scattering length $\ell$,
\begin{equation}\label{eq:t-coh}
    T_\text{coh}:\quad \xi(T_\text{coh}) \sim \ell.
\end{equation}
We will establish this linear relation, Eq.~\eqref{eq:t-coh}, in Sec.~\ref{sec:coh-peak}.
In systems with weak background scattering,
the length scales $\xi_\text{BCS}$ and $\ell$ are typically well separated with $\ell\gg\xi_\text{BCS}$.
It is hence reasonable to expect the characteristic temperatures $T_\text{mPG}$ and $T_\text{coh}$ are independent of each other.
Moreover, we introduce the temperature scale $T_m$ concerning the dip of $1/T_1T$, as marked in Fig.~\ref{fig:fig2}(a),
and investigate its evolution to elucidate the conditions under which the coherent enhancement of $1/T_1T$ is visible
upon incorporating both the bubble contribution and vertex correction.

Finally, when $T<T_c$ in Fig.~\ref{fig:fig2}(a)(b),
the system enters the SC state where the long-ranged SC correlation has been fully established.
In this regime, $1/T_1T$ is suppressed exponentially with respect to temperature,
which is expected for a gapped system.

In conclusion, the picture described above offers a universal perspective on
the NMR-$T_1$ response in 2D phase-fluctuating superconductors,
highlighting the emergent magnetic pseudogap and the coherent enhancement of $1/T_1T$ unique to $s$-wave SCs.
It is particularly significant that all these phenomena are interpreted in terms of
competing characteristic length scales within the system,
which we summarize visually in Fig.~\ref{fig:fig2}(e)(f).

\subsection{Bubble contribution: emergent scale of magnetic pseudogap}\label{sec:mpg-scale}
\begin{figure}[htbp]
    \centering\hspace*{-.5cm}
    \includegraphics[width=.98\columnwidth]{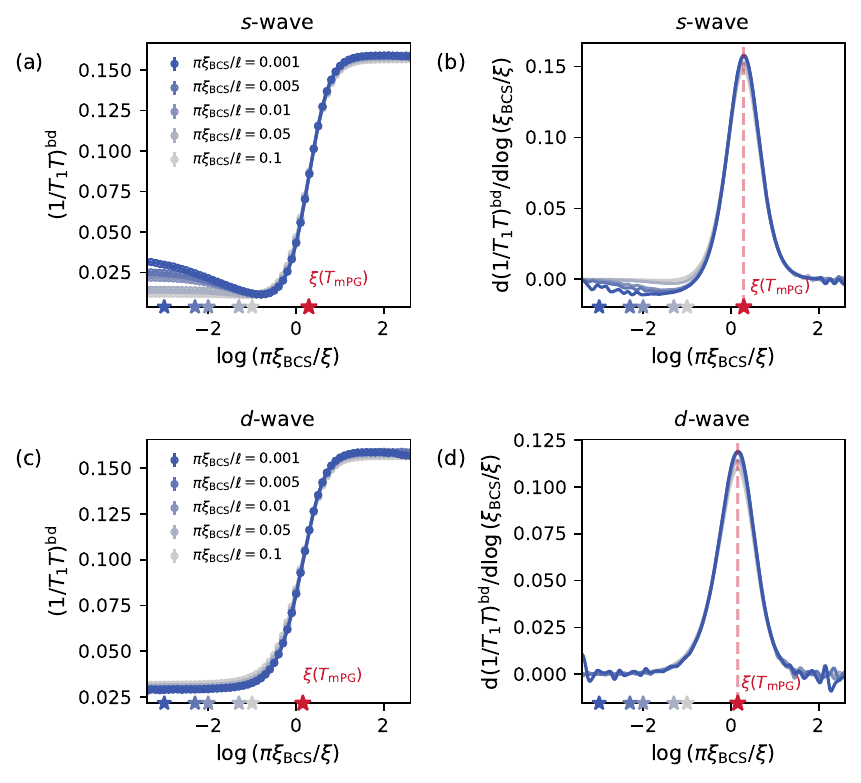}
    \caption{%
        (a)(c) Bubble-graph contribution $(1/T_1T)^\text{bd}$ with respect to the BKT correlation length $\xi$.
               We have set $E_F=2$, $\Delta_0/E_F=1/4$, $\beta\abs{\Delta_0}=5$, and $v_F=2$.
        (b)(d) Derivatives of $(1/T_1T)^\text{bd}$ in (a)(c) with respect to $\log(\xi_\text{BCS}/\xi)$.
               $\xi(T_\text{mPG})$ is identified by locating the maximum of the derivative.
               We also mark the varying $\pi\xi_\text{BCS}/\ell$ on the axis for comparison.
    }
    \label{fig:fig4}
\end{figure}

\begin{figure}[htbp]
    \centering\hspace*{-1cm}
    \includegraphics[width=.55\columnwidth]{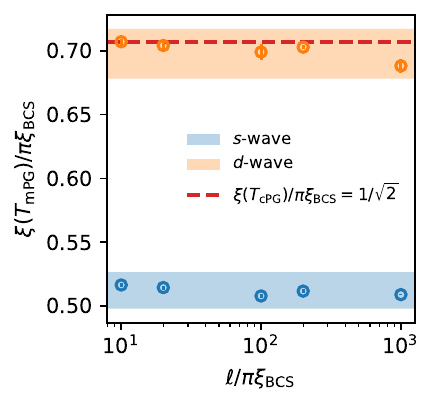}
    \caption{%
        Emergent scale of magnetic pseudogap $\xi(T_\text{mPG})$ versus background scattering length $\ell$,
        extracted from the data in Fig.~\ref{fig:fig4}.
    }
    \label{fig:fig5}
\end{figure}

\begin{figure}[htbp]
    \centering\hspace*{-.7cm}
    \includegraphics[width=.75\columnwidth]{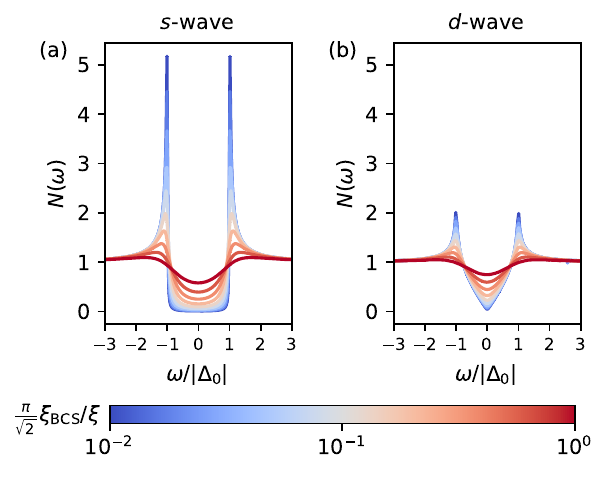}
    \caption{%
        Local density of states $N(\omega)=\sum_k A(k,\omega)$ versus $\xi_\text{BCS}/\xi$,
        revealing the onset of the charge pseudogap and coherence peak.
        The background scattering strength is set to $\Gamma_0/\abs{\Delta_0}=10^{-2}$.
    }
    \label{fig:fig6}
\end{figure}

In Fig.~\ref{fig:fig4}, we show the evolution of bubble-graph contribution to $1/T_1T$
with respect to the BKT correlation length $\xi$.
The effects of finite scattering length scale $\ell$ are examined,
while the temperature $T$ is fixed in order to exclude the trivial thermal activation effect.
As suggested in Eq.~\eqref{eq:nmr-bd},
the bubble contribution of $1/T_1T$ incorporates the electronic local density of states,
which means the evolution of magnetic pseudogap shall be closely related to the onset of charge pseudogap.
We demonstrate the charge pseudogap in Fig.~\ref{fig:fig6},
which accounts for the rapid suppression of $(1/T_1T)^\text{bd}$ in Fig.~\ref{fig:fig4}(a)(c).
In order to characterize the emergent scale of magnetic pseudogap,
we show the derivatives of $(1/T_1T)^\text{bd}$ with respect to $\log(\xi_\text{BCS}/\xi)$ in Fig.~\ref{fig:fig4}(b)(d),
and mark $\xi(T_\text{mPG})$ as the length scale where $(1/T_1T)^\text{bd}$ is most rapidly suppressed with maximal derivative.
Similar to the charge pseudogap scale given in Eq.~\eqref{eq:cpg-scale},
the magnetic pseudogap scale is determined through the ratio of BKT correlation length and BCS coherence length.
More importantly, the characteristic scale $\xi(T_\text{mPG})$
exhibits nearly no dependence on the background scattering length $\ell$ as shown in Fig.~\ref{fig:fig5}.
We remark that there is no formation of spin gap or spin fluctuations in our setup,
and $T_\text{mPG}$ emerges from pure phase fluctuations.
In addition, $T_\text{mPG}$ remains unaffected by the background scattering length $\ell$,
although $\ell$ formally regulates the divergence of $1/T_1T$ at $T_c$.

We also note that qualitative discrepancies in $(1/T_1T)^\text{bd}$ between $s$-wave and $d$-wave systems
appear when sufficiently large correlation length is reached near $T_c$.
As previously mentioned, as $\xi$ and $\ell$ approach infinity,
the infinitely sharp coherence peak manifesting in the LDOS, shown in Fig.~\ref{fig:fig6},
leads to the divergence of $(1/T_1T)^\text{bd}$.
A finite background scattering length $\ell$ broadens the coherence peak
and an enhancement of $(1/T_1T)^\text{bd}$ is possible for $s$-wave SC with reasonable $\ell$ as $\xi\to\infty$.
By weakening the background scatterings or tuning up $\ell$, the coherence effect is strengthened and so is the enhancement of $1/T_1T$.
The coherence effect becomes even more prominent when the vertex correction is taken into account for $s$-wave SC.
In contrast, the nodal structure of $d$-wave systems makes the $d$-wave coherence peak much more fragile,
and consequently the coherent enhancement is not observed in Fig.~\ref{fig:fig4}(c)
for background scattering strength as weak as $\pi\xi_\text{BCS}/\ell=\Gamma_0/\abs{\Delta_0}\simeq10^{-3}$.

\subsection{Role of vertex correction: coherent enhancement of $\bm{1/T_1T}$}\label{sec:coh-peak}
\begin{figure}[htbp]
    \centering\hspace*{-.7cm}
    \includegraphics[width=.95\columnwidth]{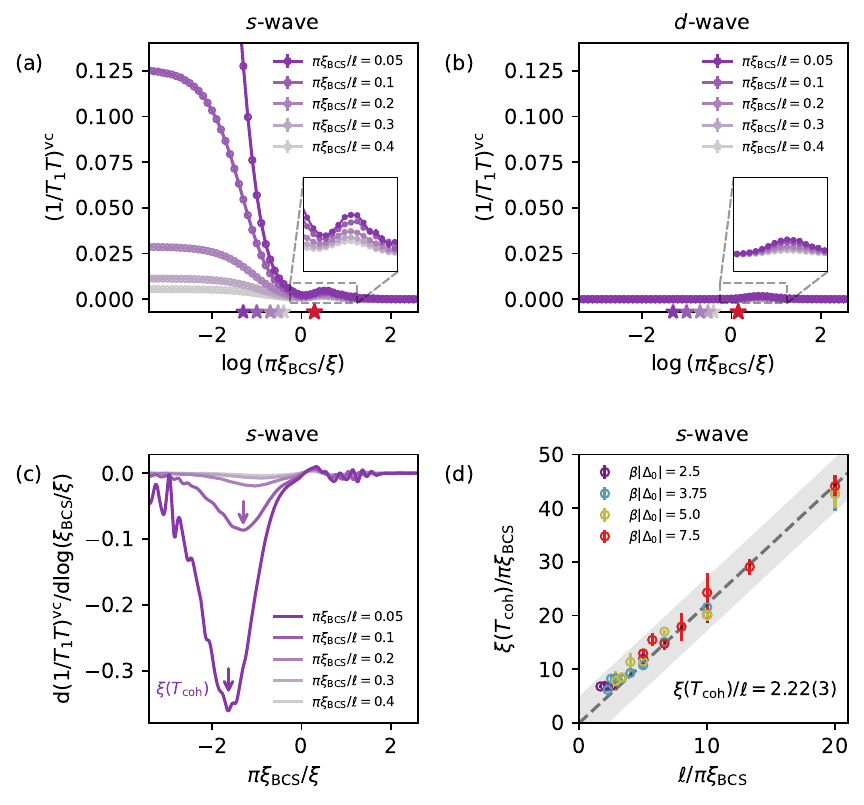}
    \caption{%
        (a)(b) Leading-order vertex contribution $(1/T_1T)^\text{vc}$ with respect to the BKT correlation length $\xi$.
               We have set $E_F=2$, $\Delta_0/E_F=1/4$, $\beta\abs{\Delta_0}=5$, and $v_F=2$.
               The magnetic pseudogap scale $\xi(T_\text{mPG})$ (in red)
               as well as the varying $\pi\xi_\text{BCS}/\ell$ (from purple to grey)
               are indicated on the axis for reference.
        (c) Derivatives of $(1/T_1T)^\text{vc}$ in (a) with respect to $\log(\xi_\text{BCS}/\xi)$ for $s$-wave system.
            The coherence scale $\xi(T_\text{coh})$ is determined as
            the scale where the derivative attains its maximum amplitude.
        (d) Linear scaling between $\xi(T_\text{coh})$ and $\ell$ for varying values of $\beta\abs{\Delta_0}$.
    }
    \label{fig:fig7}
\end{figure}

In this section, we discuss the role of vertex correction.
The vertex correction reinforces the coherence effect
since it has a higher divergence rate at $T_c$ than the bubble graph.
It is shown in Fig.~\ref{fig:fig7}(a)(b) the results of $(1/T_1T)^\text{vc}$
as a function of $\pi\xi_\text{BCS}/\xi$ with fixed $T$.
In the weak-fluctuating regime, the $s$-wave vertex correction, Fig.~\ref{fig:fig7}(a),
blows up as the background scattering strength is lowered,
while the $d$-wave vertex correction, Fig.~\ref{fig:fig7}(b), remains exactly zero
(since in Eqs.~\eqref{eq:nmr-vc}\eqref{eq:I} $I_{\sigma\bar{\sigma}}=0$ as $\xi\to\infty$).
For background scattering strength of order $\pi\xi_\text{BCS}/\ell=\Gamma_0/\Delta_0\simeq 10^{-1}$,
$(1/T_1T)^\text{vc}$ at $\xi\to\infty$ almost approaches the FL value of the bubble contribution.
We conclude that the vertex correction gives rise to a more pronounced coherent enhancement,
and makes it more likely to observe the coherence peak of $1/T_1T$
in $s$-wave SC systems even in the presence of relatively strong background scattering $\Gamma_0$.

We calculate in Fig.~\ref{fig:fig7}(c) the derivative of $s$-wave $(1/T_1T)^\text{vc}$ with respect to $\log(\xi_\text{BCS}/\xi)$,
and identify the coherence scale $\xi(T_\text{coh})$ as where the derivative attains its maximum in the amplitude,
corresponding to the location of the most rapid enhancement.
$\xi(T_\text{coh})$ is found to scale linearly with the background scattering length $\ell$
and a linear regression in Fig.~\ref{fig:fig7}(d) yields the estimation $\xi(T_\text{coh})/\ell=2.22(3)$.
Moreover, the coherent enhancement is sensitive to the dimensionless ratio $\beta\abs{\Delta_0}$,
as indicated by the Fermi-Dirac distribution in Eq.~\eqref{eq:nmr-vc}.
While varying $\beta\abs{\Delta_0}$ modifies the saturated value of $(1/T_1T)^\text{vc}$ at $\xi\to\infty$
it does not affect the rate of enhancement.
Consequently, the extracted $\xi(T_\text{coh})$ and $\ell$ exhibit the same linear relationship for varying $\beta\abs{\Delta_0}$.

Besides, close to the magnetic pseudogap scale $\pi\xi_\text{BCS}/\xi\simeq 1$,
the vertex corrections show another, yet weak, peak feature for both $s$-wave and $d$-wave pairing.
In this regime, the vertex correction, which is of order $10^{-3}$,
is two orders of magnitude smaller than the bubble contribution which is of order $10^{-1}$.
Hence, our analysis of the magnetic pseudogap scale, which relies exclusively on the bubble contribution, remains valid
since the vertex correction is negligible at this scale.
Finally in the FL limit with short correlation length $\xi/\xi_\text{BCS}\ll1$,
both $s$-wave and $d$-wave vertex corrections tend to zero and become irrelevant.
Therefore, we conclude that for temperatures near $T_c$ in $s$-wave systems,
the vertex correction is crucial as it is responsible for driving the coherent enhancement.
However, for $T$ far above $T_c$, it becomes irrelevant.
For $d$-wave pairing, the vertex correction remains insignificant
across all temperatures in the normal state and for reasonable background scattering strength.

\begin{figure}[htbp]
    \centering\hspace*{-.5cm}
    \includegraphics[width=\columnwidth]{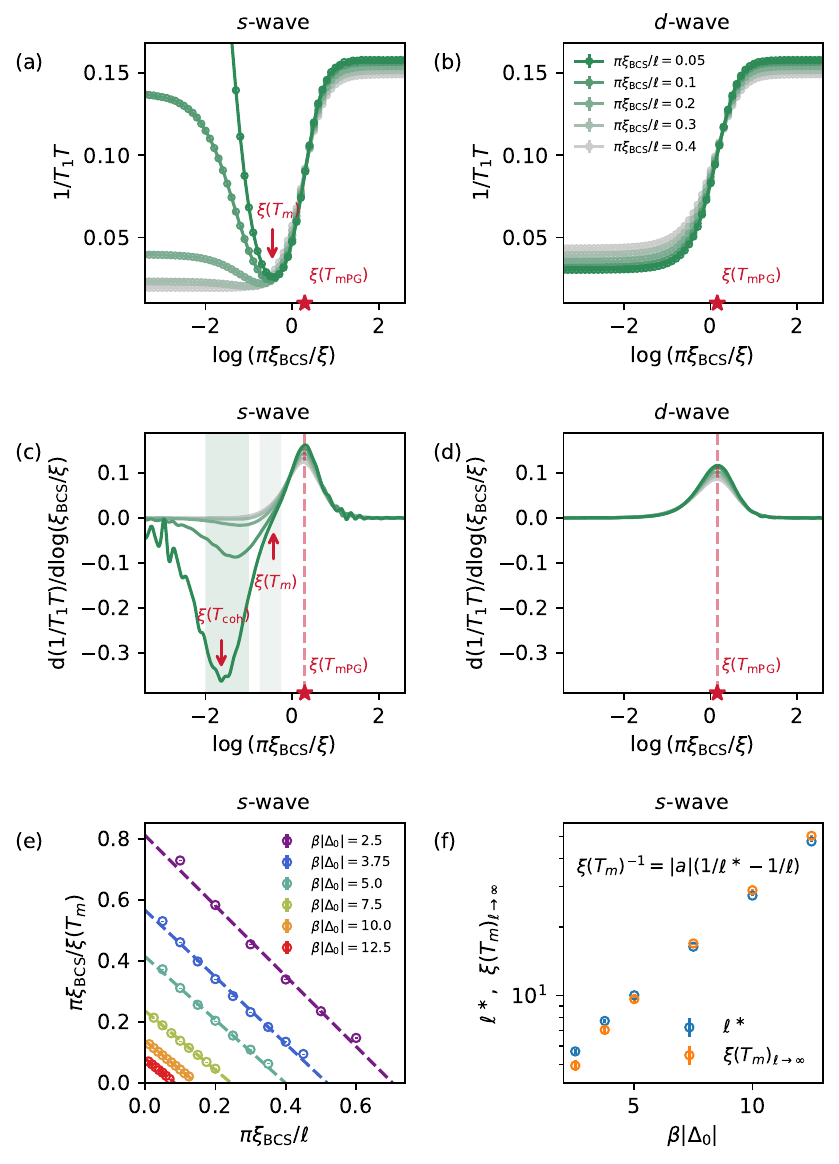}
    \caption{%
        (a)(b) $1/T_1T$ combining the bubble-graph contribution in Fig.~\ref{fig:fig4} and vertex correction in Fig.~\ref{fig:fig7}.
        (c)(d) Derivatives of $1/T_1T$ in (a)(b) with respect to $\log(\xi_\text{BCS}/\xi)$.
        (e) Characteristic scale $\xi(T_m)$ concerning the dip of $1/T_1T$ in $s$-wave system.
            We fit $\xi(T_m)^{-1}$ versus $\ell^{-1}$ via $\xi(T_m)^{-1}=\abs{a}({\ell^\ast}^{-1}-\ell^{-1})$.
        (f) Extracted critical background scattering length $\ell^\ast$
            and $\xi(T_m)_{\ell\to\infty}=\ell^\ast/\abs{a}$
            as a function of $\beta\abs{\Delta_0}$, shown on the logarithmic scale.
    }
    \label{fig:fig8}
\end{figure}

In addition, it is intriguing to figure out under what circumstance the coherent enhancement is visible in $s$-wave systems
after we combine the bubble contribution and vertex correction in Fig.~\ref{fig:fig8}.
Naively, there should be a critical background scattering length $\ell^\ast$
(which is shown to be highly sensitive to $\beta\abs{\Delta_0}$).
We define $\xi(T_m)$ as the characteristic scale concerning the dip of $1/T_1T$ in Fig.~\ref{fig:fig8}(a).
The scaling behavior of $\xi(T_m)$ with respect to $\ell$ is shown in Fig.~\ref{fig:fig8}(e) with varying $\beta\abs{\Delta_0}$.
Remarkably, $\xi(T_m)^{-1}$ exhibits an approximate linear increase
with decreasing $\ell^{-1}$ (weaker background scatterings).
It is consistent with the common understanding that
strong background scatterings $\ell<\ell^\ast$ smear out the coherence peak and lead to a vanishing $\xi(T_m)^{-1}$.
Also note that even in the clean limit $\ell^{-1}\ll1$,
$\xi(T_m)$ remains well-defined, despite a formal divergence in $1/T_1T$ in this regime.
Therefore, $\xi(T_m)$ represents another emergent scale in the normal state of $s$-wave systems,
which arises from the combined effect of SC correlation and background scatterings.
We find in Fig.~\ref{fig:fig8}(e)(f) that
the curves of $\xi(T_m)^{-1}$ versus $\ell^{-1}$ are rather sensitive to the value of $\beta\abs{\Delta_0}$.
Recall that $1/T_1T$ in Eqs.~\eqref{eq:nmr-bd}\eqref{eq:nmr-vc} involve $\frac{\mathrm{d}}{\mathrm{d}\varepsilon}n_F$
and hence the integrand at the gap edge is exponentially suppressed by $\beta\abs{\Delta_0}$.
As a consequence, with fixed scattering length $\ell$,
$1/T_1T$ at $\xi\to\infty$ grows exponentially as $\beta\abs{\Delta_0}$ diminishes.
It is thus explained that $\ell^\ast$ and $\xi(T_m)$ at $\ell\to\infty$ exhibit a positive, exponential scaling
with respect to $\beta\abs{\Delta_0}$ in Fig.~\ref{fig:fig8}(f).
These facts further suggest that it is more likely to observe the coherence peak in $1/T_1T$ for systems with smaller values of $\beta_c\abs{\Delta_0}$,
which host exponentially shorter $\ell^\ast$ such that the coherence peak can be detected if $\ell\gtrsim\ell^\ast$.

\section{Conclusion}\label{sec:conclusion}
We investigated the spin-lattice relaxation rate $1/T_1T$
for 2D phase-fluctuating superconductors with $s$-wave and $d$-wave pairing.
The effects of static phase fluctuations are taken into account within perturbation theory,
characterized by a short-ranged correlation length $\xi(T)$ obeying the BKT scaling.
Both the bubble graph and leading-order vertex correction contributing to $1/T_1T$ are then considered.
The most remarkable finding is that
the emergent magnetic pseudogap is identified arising from pure phase fluctuations without competing orders.
The associated temperature scale $T_\text{mPG}$ is determined through the comparison of two characteristic scales,
BKT correlation length $\xi(T)$ and BCS coherence length $\xi_\text{BCS}$,
which characterizes the spatial size of Cooper pair.
For the $d$-wave system, the leading-order vertex correction remains irrelevant for all temperatures in the normal state.
In contrast, the $s$-wave vertex correction becomes dominant when $T_c$ is approached from above in the normal state,
leading to a pronounced coherent enhancement of $1/T_1T$ near $T_c$.
The divergence of vertex correction at $T_c$ is closely related to
the sharp coherence peak manifesting in the electronic spectrum,
and background scatterings are vital in regulating this divergence.
As a result, the coherence scale $T_\text{coh}$, which characterizes the coherent enhancement of $1/T_1T$,
is determined by the competition between $\xi(T)$ and the background scattering length $\ell$.
In addition, we examine the evolution of scale $\xi(T_m)$ associated with the dip of $1/T_1T$ in the normal state.

Below we remark some experimental reports relevant to our theoretical findings.
First of all, the cuprate family of high-$T_c$ superconductors are known to host $d$-wave pairing,
and numerous $1/T_1$ data were provided in the literature.
For example, profound pseudogap-like behavior was revealed
for Bi2212 through NMR measurements in Ref.~\cite{ishida1998pseudogap},
and several pseudogap temperature scales $T_K^\ast$ and $T_\text{mK}$ were deduced from the Knight shift data.
In contrast, isotropic $s$-wave superconductors in quasi 2D are somehow rare.
We bring special attention to Ref.~\cite{kang2020preformed} that
magnetic pseudogap behavior was recently reported by measuring the Knight shift and $1/T_1$
in layered FeSe-based superconductors,
where organic ions are intercalated between FeSe layers to enhance the effective two-dimensionality.
Furthermore, the BKT transition and existence of preformed pairs were confirmed in these materials,
and the authors concluded that
the pseudogap behavior observed in NMR measurements is related to strong phase fluctuations.
It is thus inspiring to see that
the $1/T_1T$ data presented in their paper exhibit quite similar behavior as predicted in our theory.
However, considering the fact that
nodal superconductivity has been well established in pure FeSe films~\cite{song2011direct},
the $1/T_1T$ data in Ref.~\cite{kang2020preformed} did not show signals of coherent enhancement in the normal state.
This absence is consistent with the anticipated anisotropic superconducting state in the material.
Additionally, another relevant experimental study~\cite{li2025preformed},
employing combined NMR and Nernst measurements,
suggested that the precursor of spin resonance observed in a quasi-2D triclinic iron pnictide superconductor
is highly related to the preformed Cooper pairs driven by phase fluctuations,
which thereby supports our core findings.

In a very recent preprint~\cite{yang2025spinlattice},
the spin-lattice relaxation rates of $s$-wave and $d$-wave BKT superconductors were examined
through conducting classical Monte Carlo samplings over fluctuating pairing configurations generated by the XY model.
The authors reported a Hebel-Slichter coherence peak in the normal state of $s$-wave SC,
which they attributed to the singular behavior of quasiparticle LDOS as also suggested in our work.
In contrast, $d$-wave SC showed no Hebel-Slichter coherence peak;
instead, the suppression of $1/T_1T$ in the normal state of $d$-wave system coincided with the development of charge pseudogap.
All their results are consistent with our theoretical predictions,
therefore providing convincing numerical support to our findings.
Moreover, for further numerical verifications among correlated systems,
the attractive-$U$ Hubbard model is particularly notable,
as it has long been confirmed~\cite{moreo1991two-dimensional,moreo1992quasiparticle}
to host an $s$-wave SC state at finite doping
and undergo a BKT superconducting transition at finite temperature.
Since the model can be exactly simulated with sign-problem-free quantum Monte Carlo method,
it serves as a promising platform for testing our predictions on
the emergent magnetic pseudogap and coherent enhancement of $1/T_1T$.
Further investigations along this line are awaited.

Lastly, there is a broad consensus that
spin correlations are pervasive and crucial in high-$T_c$ superconductivity.
For BCS superconductors, it has been shown that
the celebrated Hebel-Slichter peak can be suppressed
in the presence of strong antiferromagnetic fluctuations~\cite{cavanagh2021fate}.
It is hence plausible to argue that the scatterings mediated by fluctuating spin correlations
may also lead to the fading of $1/T_1T$ coherence peak predicted in this work for $s$-wave systems.
Whether the emergent scale of magnetic pseudogap
will be altered fundamentally by spin fluctuations remains an open question.
Therefore, an in-depth examination on the interplay between superconducting and spin correlations
shall help extend our understanding of magnetic pseudogap phenomenon to more complex and realistic high-$T_c$ materials.

\begin{acknowledgments}
We thank Kai Sun, Tao Shi and Yi Zhou for invaluable discussions.
This work is supported by
the National Key R\&D Program of China (Grant No.~2022YFA1403402),
the National Natural Science Foundation of China (Grant No.~12174068),
the Science and Technology Commission of Shanghai Municipality (Grant Nos.~24LZ1400100 and 23JC1400600),
and the Shuguang Program of Shanghai Education Development Foundation and Shanghai Municipal Education Commission.
The high-dimensional integrals arising in this work are evaluated using the open-source Python package
\href{https://github.com/gplepage/vegas}{vegas},
which implements high-performance adaptive multidimensional Monte Carlo integration.
The authors also acknowledge \href{https://www.paratera.com}{Beijing PARATERA Tech Co., Ltd.}
for providing the computational resources used in this work.
\end{acknowledgments}

\bibliographystyle{apsrev4-2}
\bibliography{ref.bib}

\end{document}